\documentclass[9pt, conference]{IEEEtran}
\IEEEoverridecommandlockouts
\usepackage{cite}
\usepackage{amsmath,amssymb,amsfonts}
\usepackage{algorithmic}
\usepackage{graphicx}
\usepackage{textcomp}
\usepackage{xcolor}
\usepackage{url}
\usepackage{hyperref}
\hypersetup{
    colorlinks=true,
    linkcolor=blue,
    filecolor=magenta,      
    urlcolor=cyan,
    pdftitle={Overleaf Example},
    pdfpagemode=FullScreen,
    }

\def\BibTeX{{\rm B\kern-.05em{\sc i\kern-.025em b}\kern-.08em
    T\kern-.1667em\lower.7ex\hbox{E}\kern-.125emX}}
\begin{document}

\title {ViolinDiff: Enhancing Expressive Violin Synthesis with \\Pitch Bend Conditioning

}

\author{\IEEEauthorblockN{Daewoong Kim}
\IEEEauthorblockA{\textit{Department of Artificial Intelligence} \\
\textit{Sogang University}\\
Seoul, Korea \\
\texttt{nueyet@sogang.ac.kr}}
\and
\IEEEauthorblockN{Hao-Wen Dong}
\IEEEauthorblockA{\textit{Department of Performing Arts Technology} \\
\textit{University of Michigan}\\
Ann Arbor, MI, USA \\
\texttt{hwdong@umich.edu}}
\and
\IEEEauthorblockN{Dasaem Jeong}
\IEEEauthorblockA{\textit{Department of Art \& Technology} \\
\textit{Sogang University}\\
Seoul, Korea \\
\texttt{dasaemj@sogang.ac.kr}}
}

\maketitle

\begin{abstract}
Modeling the natural contour of fundamental frequency (F0) plays a critical role in music audio synthesis. However, transcribing and managing multiple F0 contours in polyphonic music is challenging, and explicit F0 contour modeling has not yet been explored for polyphonic instrumental synthesis. In this paper, we present ViolinDiff, a two-stage diffusion-based synthesis framework. For a given violin MIDI file, the first stage estimates the F0 contour as pitch bend information, and the second stage generates mel spectrogram incorporating these expressive details. The quantitative metrics and listening test results show that the proposed model generates more realistic violin sounds than the model without explicit pitch bend modeling.
Audio samples are available online: \url{daewoung.github.io/ViolinDiff-Demo}.
\end{abstract}

\begin{IEEEkeywords}
Violin Synthesis, Neural Audio Synthesis, Pitch Bend Modeling, Expressive Performance, Diffusion Models
\end{IEEEkeywords}

\section{Introduction}
Recent advances in neural audio synthesis have yielded impressive results across various domains, such as speech~\cite{glowtts, gradtts, vits}, singing voice~\cite{xiaoicesing, visinger, diffsinger, expressivesinger}, and instrumental sounds~\cite{mididdsp, deepperformer, multiinst, performerdiff, kim2024expressive}. 
However, while the synthesized result of speech or singing voice achieved a high mean opinion score (MOS) of around 3.9~\cite{diffsinger, expressivesinger, visinger}, the quality score of music instrument synthesis typically lies around 3.1~\cite{kim2024expressive}, indicating significant room for improvement.

A key challenge in instrumental music synthesis, as compared to singing voice synthesis (SVS), lies in its polyphonic nature. Instruments often play multiple notes simultaneously, whereas singing is predominantly monophonic. Many recent SVS models~\cite{visinger, diffsinger, expressivesinger} utilize explicitly modeled F0 (fundamental frequency) contours. VISinger~\cite{visinger} has demonstrated the effectiveness of this approach. It predicts F0 from note pitch, note duration, and lyrics, which significantly enhances the naturalness of the synthesized output.
However, this approach becomes problematic for polyphonic music, where multiple notes can exist in a single time frame, making it infeasible to encode F0 contours as a simple 1D sequence.
Furthermore, estimating F0 for polyphonic audio is substantially more complex than for monophonic audio due to the overlapping frequencies and interactions between simultaneous notes, which adds another layer of difficulty in acquiring training data for instrument synthesis with F0 modeling.

In this paper, we introduce \textbf{ViolinDiff}, a diffusion-based violin synthesis framework that addresses these limitations by modeling pitch bend information of MIDI and incorporating it as an explicit condition of audio synthesis.
We trained the model using a dataset released along with a recent violin transcription model~\cite{tamer2023high} that includes detailed polyphonic pitch deviation as MIDI pitch bend . 
We evaluate the effectiveness of ViolinDiff using quantitative metrics 
and listening tests, demonstrating the advantages of incorporating pitch bend information in violin synthesis.
Our contribution lies in \textit{i)} proposing a bend roll, a new encoding scheme for polyphonic F0 contour, \textit{ii)} a novel evaluation metric for vibrato prediction, and \textit{iii)} implementing a high-quality violin solo synthesis model with a recently published dataset made by automatic transcription and alignment.

\section{Related Works}
Several studies have been conducted on instrument synthesis, including research that encompasses violin sound. MIDI-DDSP~\cite{mididdsp} generates expressive details such as vibratos and attack noise directly from MIDI data, improving the realism of synthesized audio. However, MIDI-DDSP is primarily focused on monophonic MIDI, limiting its ability to handle complex polyphonic music. Deep Performer \cite{deepperformer}, a transformer-based model, synthesizes polyphonic audio from musical scores but lacks fine-grained pitch control.

Recent advancements in diffusion-based music synthesis models \cite{multiinst, performerdiff} have improved audio fidelity and expressiveness. Hawthorne et al. \cite{multiinst} proposed a T5-based architecture for synthesizing multi-instrument audio. Maman et al. \cite{performerdiff} incorporated performer embeddings to control timbre and style for orchestral instruments and the acoustic condition of each recording. These performer embeddings helped enhance the overall quality of synthesized audio. However, these models do not explicitly model F0 contours, 
which leaves a room to be improved in synthesis quality and controllability.

{ 
\begin{figure*}[t]
\centering
\includegraphics[height=0.25\textheight, keepaspectratio]{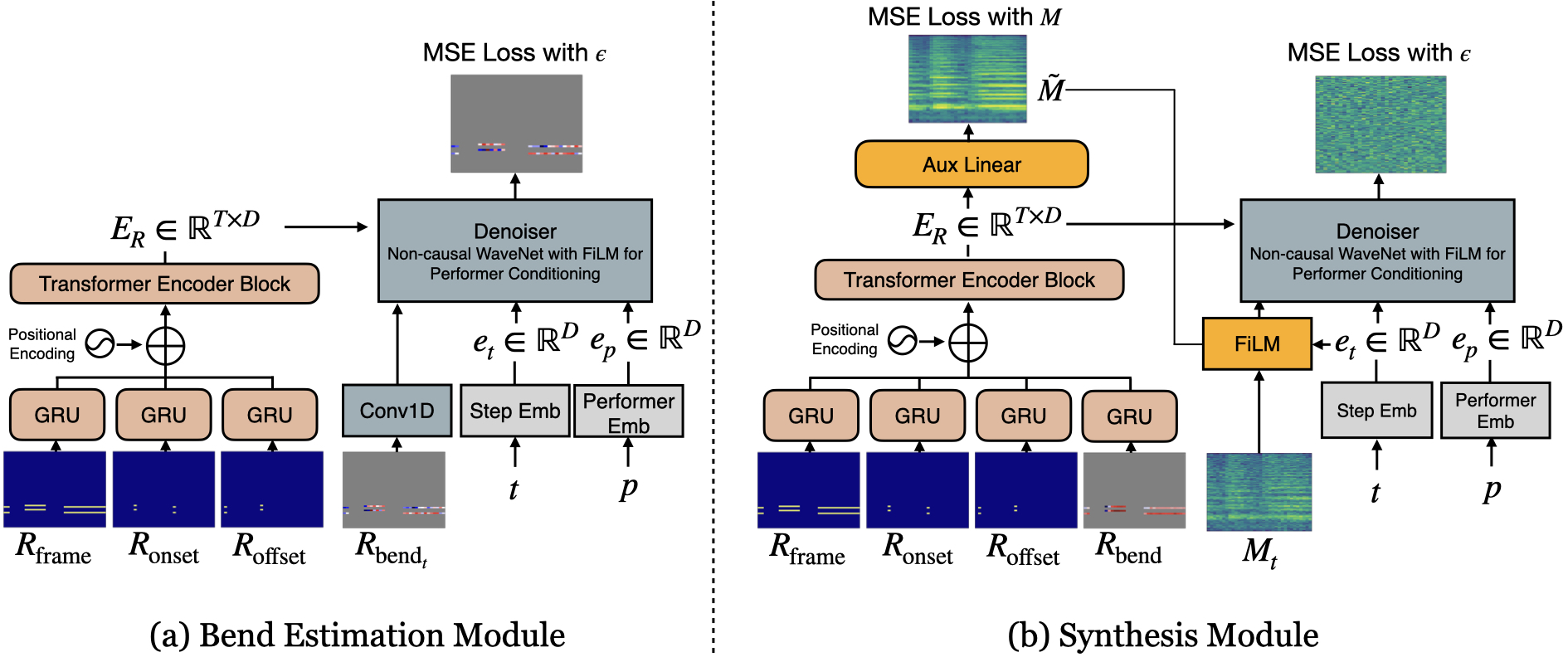}
\caption{  Overview of the architectures for (a) Bend Estimation Module and (b) Synthesis Module. \( R_{\text{bend}_t} \) and \( M_t \) represent the \( R_{\text{bend}} \) and the \( M \) with noise applied at step \( t \) of the diffusion process. 
The Denoiser in both modules adopts the non-causal WaveNet architecture as employed in DiffSinger \cite{diffsinger}, with additional modifications to incorporate performer conditioning through FiLM layers, where each residual block is conditioned on performer embeddings.}
\label{fig:1}
\end{figure*}
}
\section{Dataset}
We utilized the dataset introduced in \cite{tamer2023high}, which was generated through an iterative learning and alignment process using weak labels, pairing violin recordings with their corresponding publicly available scores. The dataset comprises 34 hours of solo violin performances, 60 etudes by Wohlfahrt (Op.~45), 36 etudes by Kayser (Op.~20), and 24 caprices by Paganini (Op.~1). These pieces were performed by multiple violinists, resulting in 1,021 recordings from 22 distinct performers. The dataset includes both audio recordings and synchronized MIDI files, with the MIDI files providing crucial multi-pitch bends information. 

After excluding unavailable recordings due to removed YouTube links and eliminating one performer due to poor audio quality, we used a total of 961 audio recordings from 21 distinct performers. For the validation set, we selected 8 recordings of Kayser Op.~20 No.~16. The test set includes a diverse range of pieces—Kayser Op.~20 Nos. 1, 18, and 36, Paganini Op.~1 Nos.~5, 13, and 24, and Wohlfahrt Op. 45 Nos. 1, 30, and 60—comprising 78 audio recordings across 9 pieces. This setup enabled a comprehensive evaluation of the model's performance across various compositions and performance styles.

\section{Method}
In this section, we introduce the structure of ViolinDiff, which consists of a bend estimation module and a mel spectrogram synthesis module, both in forms of denoising diffusion probablistic models~\cite{ddpm}.

To encode MIDI input, we employed piano roll representation as it enables encoding polyphonic pitch while being aligned to the mel spectrogram at frame level. If a mel spectrogram $M \in \mathbb{R}^{F\times T}$ consists of $T$ time frames and $F$ mel bins, the input piano roll is a 2D binary matrix $R \in \{0, 1\}^{P \times T}$, where $P$ denotes the number of pitches. To emphasize the importance of the onset and offset of each note along with its frame-wise activation, we employed three different types of rolls: $R_{\text{frame}}, R_{\text{onset}}, R_{\text{offset}}$. We followed the definition of each roll from piano music transcription research~\cite{hawthorne2018}.

In addition to these multiple binary piano rolls, we also include pitch bend information in the model's input. Pitch bends in MIDI are discrete events, each with a pitch bend value and a corresponding time, which enable the continuous change of pitch. These events often occur at finer intervals than what traditional piano roll representations can capture. Therefore, we calculate frame-wise pitch bend value by weighted average considering the duration of pitch bend events within each frame, which results in $R_{\text{bend}} \in (-1, 1)^{P \times T}$.

The total input of the synthesis module can be represented as \( X = \{M_t, \mathcal{R}, t, p\} \), where $M_t$ is mel spectrogram with $t$-step gaussian noise, $\mathcal{R} = \{R_{\text{frame}}, R_{\text{onset}}, R_{\text{offset}}, R_{\text{bend}}\}$ denotes input MIDI information encoded in piano roll formats that are aligned in frame-level with the mel spectrogram, $t \in {0, 1, \dots, N}$ for diffusion time step, and $p$ for performer ID. $N$ denotes the number of diffusion steps.

Bend estimation module adjusts this structure by replacing the mel spectrogram with the direct bend roll \( R_{\text{bend}} \), using \( X = \{R_{\text{bend}, t},\mathcal{R}_{\text{w/o bend}}, t, p\} \), where \( \mathcal{R}_{\text{w/o bend}} = \{R_{\text{frame}}, R_{\text{onset}}, R_{\text{offset}}\} \).

\subsection{Synthesis Module}\label{AA}

In the synthesis module, the encoder processes the rolls $\mathcal{R}$ as input, through a GRU layer, followed by Transformer Encoder Blocks~\cite{transformer}, producing the note feature sequence \( E_R \in \mathbb{R}^{T\times D} \) as presented in Fig. ~\ref{fig:1}~(b), where $D$ denotes model dimension size. A simple linear layer then generates an auxiliary mel spectrogram  \( \tilde{M} \) from \( E_R \). For the denoiser of the synthesis module, we adopt the non-causal WaveNet\cite{wavenet} architecture as employed in DiffSinger \cite{diffsinger}, with additional modifications to incorporate performer conditioning through FiLM layers~\cite{perez2018FiLM}. To denoise the noisy mel spectrogram $M_t \in \mathbb{R}^{F\times T}$, we first add a time step embedding $e_t \in \mathbb{R}^D$ by passing it through a linear layer $f: \mathbb{R}^D \rightarrow \mathbb{R}^F $ to each time frame of $M_t$. The combined input is modulated again via a FiLM layer, using the auxiliary mel spectrogram \( \tilde{M} \) as a conditioning input.

Each residual block in the denoiser receives conditioning inputs \( e_t \), \( e_p \), and \( E_R \). The performer embedding \( e_p \in \mathbb{R}^D \) is processed through FiLM layers to modulate the internal processing within the residual block, while the time step embedding \( e_t \) and the note features \( E_R \) are added directly, following the approach of \cite{diffsinger}.

The model is trained to simultaneously minimize two L2 losses: the first, following the DDPM \cite{ddpm} loss formulation, predicts the noise (\(\epsilon\)) added to \(M_t\), while the second minimizes the error between the predicted auxiliary mel spectrogram \(\tilde{M}\) and the ground truth mel spectrogram \(M\). We did not use a weight to balance the two losses.

\begin{figure*}[ht!bp]
\centering
\includegraphics[width=0.85\textwidth, keepaspectratio]{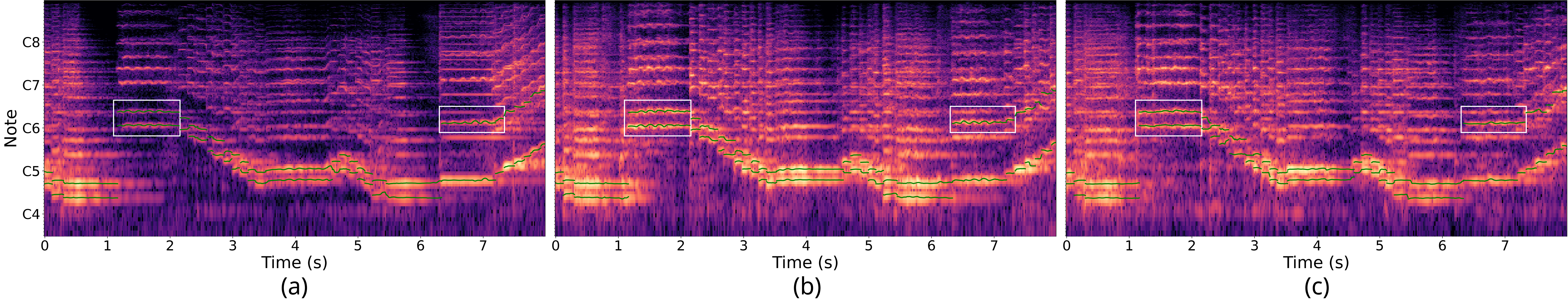}

\caption{Spectrogram comparison with F0 (green curves). From left to right: (a) Original audio, (b) ViolinDiff, and (c) Baseline \textit{NoBend} model. Spectrograms are cropped to exclude frequencies below than violin's lowest note.} 
 
\label{fig:2}
\end{figure*}

\begin{table*}
\caption{Model Performance Comparison on FAD and Vibrato Metrics}
\begin{center}
\begin{tabular}{|c|c|c|c|c|c|c|c|c|c|c|c|}
\hline
\textbf{} & \multicolumn{6}{|c|}{{FAD Metrics}} & \multicolumn{4}{|c|}{{Vibrato Metrics}} \\
\hline
{Model} & \multicolumn{2}{|c|}{{All-FAD  $\downarrow$}} & \multicolumn{2}{|c|}{{Performer-FAD  $\downarrow$}} & \multicolumn{2}{|c|}{{Piece-FAD  $\downarrow$}} & \multicolumn{2}{|c|}{{F1 Score $\uparrow$}} & \multicolumn{2}{|c|}{{Perf-MAE $\downarrow$}} \\
\cline{2-11} 
\textbf{} & {\textit{L-Audio}} & {\textit{L-Music}} & {\textit{L-Audio}} & {\textit{L-Music}} & {\textit{L-Audio}} & {\textit{L-Music}} & {\textit{TAPE}} & {\textit{MUSC}} & {\textit{TAPE}} & {\textit{MUSC}} \\
\hline
ViolinDiff  & \textbf{0.040} & \textbf{0.046}  & \textbf{0.079} & \textbf{0.083} & \textbf{0.074} & \textbf{0.078} & \textbf{0.567} & \textbf{0.465} & \textbf{0.140} & \textbf{0.229} \\
Baseline: \textit{NoBend} & 0.048 & 0.052  & 0.083 & 0.085 & 0.081 & 0.080 & 0.408 & 0.343 & 0.262 & 0.257 \\
\hline
GT Encoded & 0.028 & 0.033  & 0.062 & 0.064 & 0.025 & 0.029  & 0.746 & 0.724 & 0.075 & 0.064 \\
Ground Truth Raw & 0.013 & 0.015  & 0.046  & 0.044  & 0.000  & 0.000 & 1.0 & 1.0 & 0.000 & 0.000 \\
\hline
\end{tabular}
\label{tab1}
\end{center}
\end{table*}

\subsection{Bend Estimation Module}\label{BB}

The bend estimation module, which is trained separately, is adapted from the synthesis module to handle pitch bend data directly.
Unlike the synthesis module, which conditions its denoising process on an auxiliary mel spectrogram \( \tilde{M} \), the bend estimation module operates directly on the noisy bend roll \( R_{\text{bend}_t} \) (Fig.~\ref{fig:1}~(a)).
During the diffusion, we only add noise to the regions with note activation. Other regions were excluded in noise addition and loss calculation.

\subsection{Diffusion Process}\label{BB}
Our approach is based on the principles of Denoising Diffusion Probabilistic Models (DDPM) ~\cite{ddpm}.  This framework offers a robust probabilistic method for generating high-quality samples through a learned iterative denoising process, and its effectiveness in music synthesis has been demonstrated in various studies \cite{diffsinger, kim2024expressive, performerdiff, multiinst}.

In the synthesis module, \( N \) is set to 200, while in the bend estimation module, \( N \) is set to 100. For both modules, the \( \beta \) values increase linearly from \( \beta_1 = 10^{-4} \) to \( \beta_N = 0.06 \). Both modules use Classifier-Free Guidance (CFG)~\cite{cfg}, with condition dropout applied independently to \( e_p \), \( e_m \), and \( \tilde{M} \) at a probability of 0.1 during training. During sampling, guidance weights of 1.25 are used for the synthesis module and 3.0 for the bend estimation module.

To convert the generated mel spectrogram back to audio, we used the SoundStream Vocoder~\cite{zeghidour2021soundstream}, which has also been employed in previous works \cite{multiinst, kim2024expressive, performerdiff}.

\section{Experiments}

\subsection{Implementation Details}\label{Implementation Details} 
The audio is downsampled to 16kHz, with mel spectrogram computed using a 1024 FFT size, 640 window length, 320 hop length, and 128 Mel bins. The synthesis module is trained on 256-frame spectrograms (5.12 seconds), while bend estimation module uses 512-frame spectrograms (10.24 seconds). The piano roll is limited to 54 pitches (MIDI pitch 55 to 108). 

The synthesis module uses separate bidirectional GRU layers (128 hidden units per direction, 2 layers) for frame, onset, offset, and bend information, while the bend estimation module processes only frame, onset, and offset information. The GRU outputs are summed and fed into a transformer encoder (256 input dimension, 1024 FFN hidden size, 6 layers, 4 attention heads per layer). Both models share a denoiser with 256 residual channels, and 20 layers without dilation. Both models are trained using the Adam optimizer with a learning rate of 1e-4. The batch size is set to 128. The synthesis module is trained for 280K steps, while the bend estimation module is trained for 112K steps. Both synthesis module and bend estimation module have approximately 25M parameters each.

For generating longer audio than the training settings, we followed the methodology outlined in \cite{performerdiff}, which has proven effective for music synthesis. This approach was originally adapted from motion generation techniques \cite{motion1, motion2}. 

\subsection{Baseline Model: ViolinDiff without Pitch Bend}\label{BB}
To evaluate the advantage of modeling pitch bend explicitly, we also designed a baseline model, \textit{NoBend}, that does not use pitch bend information during the synthesis. \textit{NoBend} model only uses the synthesis module, and the module takes $\mathcal{R}_{\text{w/o bend}}$ instead of $\mathcal{R}$. Therefore, the baseline has to model the F0 trajectory implicitly during the synthesis without explicit pitch bend information.

\subsection{Fréchet Audio Distance (FAD)}\label{sec:FAD}
Fréchet Audio Distance (FAD) measures perceptual similarity by calculating the distance between multivariate Gaussian distributions fitted to embeddings of generated and reference audio \cite{fad}. Instead of the commonly used VGGish model~\cite{vggish}, we employed the L-CLAP audio and L-CLAP music models from \cite{lclap}, which have been shown to perform effectively for FAD calculation in \cite{adaptivefad}. Embeddings for FAD were extracted using the \texttt{fadtk} library \cite{adaptivefad}.

1) \textbf{All-FAD} measures general audio quality by comparing the entire dataset to the synthesized audio generated from the test set MIDI data.

2) \textbf{Performer-FAD} assesses how well the model replicates the unique characteristics (e.g., timbre, acoustics) of each performer. It is calculated by comparing the synthesized audio generated from the test set MIDI data for a specific performer %(e.g., Performer 0)
to the corresponding real audio from the entire dataset of that same performer.

3) \textbf{Piece-FAD} evaluates how well the model reproduces each piece by comparing the synthesized audio of a specific piece to the real performances of that piece by multiple performers in the test set. For example, if Kayser Op.~20 No.~16 was performed by Performer~0 and Performer~1, the model's generated audio for both performers would be compared to their real performances. This comparison helps assess how accurately the model captures the overall characteristics of the piece across different performers, rather than focusing on the distribution of arbitrary musical pieces.

\subsection{Vibrato Evaluation}\label{sec:vibrato_analysis}
To evaluate the naturalness of vibrato, one of the key expressive features in violin performances, we propose novel metrics for its usage. To measure the usage of vibrato for each note, we followed the method proposed in MIDI-DDSP \cite{mididdsp}, which applies the discrete Fourier transform (DFT) to the F0 sequence of each note to calculate the vibrato value, which is the periodicity of the F0 contour. 
F0 extraction was carried out using two different models: TAPE \cite{tape}, which is specialized for monophonic F0 extraction of violin audio, and MUSC \cite{tamer2023high}, which can handle polyphonic sections by extracting multiple F0s. Since TAPE is designed for monophonic input, we excluded polyphonic notes from the vibrato evaluation when using this model. The vibrato analysis was conducted for each note as specified in the ground truth MIDI. For the ground truth (GT) vibrato values, we used the F0s extracted from the GT audio, calculating the vibrato separately using both the TAPE and MUSC models. 
Note that we only consider notes that are longer than 0.2 seconds and vibrato rate between 3 to 9 Hz during the vibrato value calculation following~\cite{mididdsp}. 
While the vibrato value is calculated using a scalar value to show how clear the vibrato presence is, we converted it to a boolean value, i.e. absence or presence of vibrato, for the evaluation to adjust noises caused by errors in F0 estimation.

1) \textbf{F1 Score:} 
To evaluate the accuracy of vibrato generation, we compared the ground truth (GT) vibrato for each note with the vibrato derived from the F0 extracted from the model's synthesized audio. For each note, we determined whether the generated vibrato correctly matched the GT vibrato in terms of its presence or absence. We calculated the F1 Score for each individual piece and then computed the macro-average F1 Score by averaging the scores across all pieces.

2) \textbf{Perf-MAE:} 
Vibrato is a subjective element in violin performances, and its usage is a matter of stylistic choice. However, we can evaluate how well the model captures typical vibrato practices by analyzing how multiple performers apply vibrato within each note of a piece. Using a symbolic music alignment algorithm~\cite{allign}, we aligned the MIDI data across all performers for each piece, allowing us to examine and compare the vibrato usage for the same notes by different performers. 
The goal was to determine whether the model could replicate common performance practices, such as applying vibrato when most performers do. 
Only the notes sustained for at least 0.2 seconds by all performers were included in the final evaluation. For each note, we calculated the ratio of performers who played vibrato for the note and calculated the mean absolute error (MAE) of the ratio between GT and the model's predictions. 
This metric assesses whether the model's prediction on the probability of using vibrato for each note is similar to ground truth distribution.

\subsection{Listening Test}\label{listening}
The listening test aimed to evaluate the realism of synthesized violin sounds using the MUSHRA protocol \cite{mushra}, with a focus on how well the pitch expression and overall timbre matched real violin performances. 
15-second segments were randomly selected from Kayser Op. 20 No. 36, Wohlfahrt Op. 45 No. 60, and Paganini Op. 1 Caprices Nos. 5, 13, and 24. 
For Kayser and Wohlfahrt, performer embeddings were sourced from two performers who had recorded both pieces, while for Paganini, embeddings came from three renowned performers. Two segments were used for Paganini Caprices Nos. 13 and 24, while one segment was used for the other pieces. 
% A zero-shot evaluation 
To evaluate the model's generalizability on out-of-domain input, we also included three 15-second segments of Bach, Massenet, and Ysaÿe's piece from YouTube-sourced audio, with MIDI generated by using the model from \cite{tamer2023high} to transcribe and align the audio with separately obtained scores. Performer embeddings for these segments were randomly selected from the same three performers used for the Paganini Caprices. 

The models compared included those by Hawthorne et al.\footnote{Samples were generated using the published code and weights.}~\cite{multiinst}, Maman et al.\footnote{Samples were graciously provided by the authors of the original paper.}~\cite{performerdiff}, and our ViolinDiff and the baseline model, with the GM soundfont serving as the lowest anchor. The test, covering 10 segments, was conducted on the webMUSHRA platform \cite{webmushra}. Out of 34 participants, we filtered out eight who rated the reference sample above 90 in less than 80\% of the questions, and three who completed the test in less than 10 minutes. Therefore, total of 23 participants were counted for the analysis.

\begin{figure}[t]
\centering
 
\includegraphics[width=0.4\textwidth, keepaspectratio]{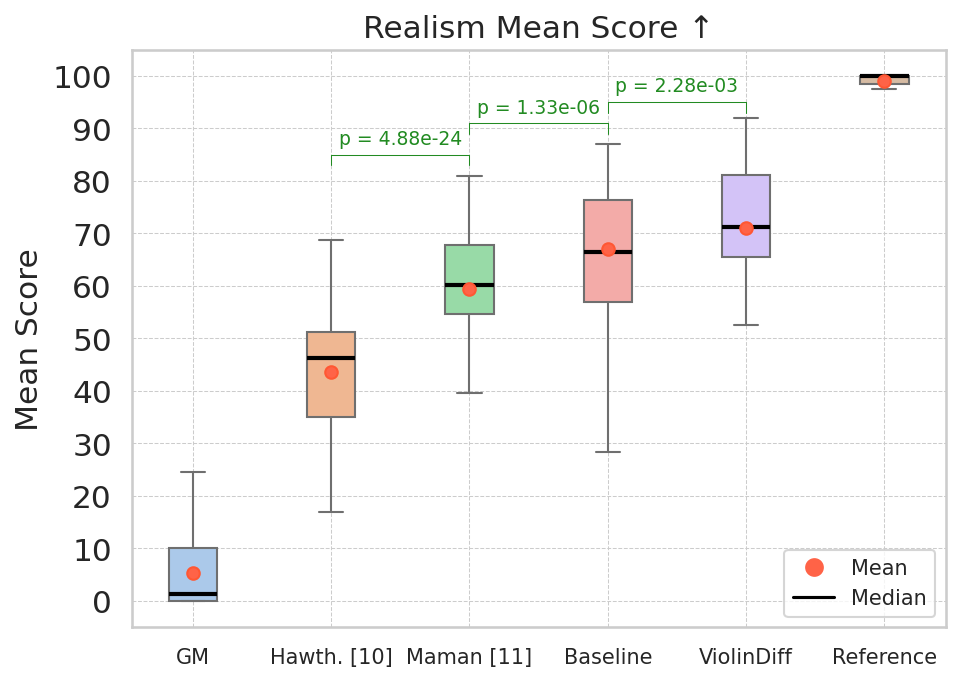}
\caption{The box plot illustrates the results of the MUSHRA realism listening test. The p-values between models were calculated using the Wilcoxon signed-rank test. The green lines connect the compared models.}

\label{fig:3}
\end{figure}

\section{Results}\label{result}

As shown in Table \ref{tab1}, ViolinDiff consistently outperformed the baseline model across all FAD metrics (All-FAD, Performer-FAD, and Piece-FAD) in both the L-audio and L-music. The lower All-FAD score indicates more globally realistic audio, while the improved Performer-FAD and Piece-FAD scores suggest better capture of performer-specific and piece-specific qualities. These results highlight the effectiveness of explicitly modeling pitch bend information in achieving expressive and realistic music synthesis.

In the vibrato evaluation, ViolinDiff outperformed the baseline model in both F1 Score and Perf-MAE metrics. 
The result shows that explicit pitch bend modeling enhances vibrato modeling compared to implicit modeling of the baseline, both in terms of personalized style and commonness across the performers. Additionally, Fig.~\ref{fig:2} shows a detailed example that ViolinDiff can model clearer vibrato in polyphonic excerpt compared to the baseline model.

The listening test results in Fig.~\ref{fig:3} show that ViolinDiff achieved the highest realism score among the synthesized models, with a mean score of 70.96, outperforming the baseline model (66.99), showing that incorporating pitch bend information enhances synthesis quality. Wilcoxon signed-rank test showed a significant difference ($p<0.003$) between these two models. 
Although it is difficult to compare the result with other models straightforwardly as the training data is different, our proposed model showed better result compared to the previous diffusion models such as Maman et al.~\cite{performerdiff} (59.47) and Hawthorne et al.~\cite{multiinst} (43.67). The samples are available \href{http://daewoung.github.io/ViolinDiff-Demo}{here}.

\section{Conclusion}
We presented ViolinDiff, a diffusion-based MIDI-to-audio synthesis model that predicts polyphonic pitch bend and leverages it to synthesize expressive violin performances. Our two-stage architecture, validated through quantitative metrics and listening tests, consistently produced more realistic sound, highlighting the effectiveness of explicit modeling pitch bend information in violin synthesis. For future work, we aim to incorporate additional control parameters such as tempo and articulation to further refine expressiveness. Additionally, extending the model to synthesize other instruments would broaden its applicability.

\newpage\newpage
\bibliographystyle{IEEEtran}
\bibliography{refs}

\end{document}